# Spatial organization and interactions of harvester ants during foraging activity

Jacob D. Davidson[1,2] and Deborah M. Gordon[3]
1. Max Planck Institute for Ornithology, Department of Collective Behavior, Konstanz, Germany
2. Princeton University, Department of Ecology and Evolutionary Biology, Princeton, USA
3. Stanford University, Department of Biology, Stanford, USA

## Abstract
Local interactions, when individuals meet, can regulate collective behavior. In a system without any central control, the rate of interaction may depend simply on how the individuals move around. But interactions could in turn influence movement; individuals might seek out interactions, or their movement in response to interaction could influence further interaction rates. We develop a general framework to address these questions, using collision theory to establish a baseline expected rate of interaction based on proximity. We test the models using data from harvester ant colonies. A colony uses feedback from interactions inside the nest to regulate foraging activity. Potential foragers leave the nest in response to interactions with returning foragers with food.  The time series of interactions and local density of ants show how density hotspots lead to interactions that are clustered in time. A correlated random walk null model describes the mixing of potential and returning foragers. A model from collision theory relates walking speed and spatial proximity with the probability of interaction. The results demonstrate that although ants do not mix homogeneously, trends in interaction patterns can be explained simply by the walking speed and local density of surrounding ants.

## Introduction

Collective behavior arises from interactions among individuals, and these interactions often occur locally. The probability of interaction thus depends on spatial patterns of movement, while in turn, the pattern of movement may be influenced by interactions [1], and on variation among individuals in their use of space [2], [3]. A central question is how the movements of individuals, and constraints from the local environment, shape the local interactions that lead to collective behavior.

The relation between motion patterns of particles and the probability of interaction is described by the scattering cross section in gas dynamics, spectroscopy, and particle physics [4]. This specifies the probability that two nearby particles interact in a collision, which alters the motion and/or energy of the particles. For example, two high-energy particles may interact in an inelastic collision that results in a new particle [5]. In a mixture of gas particles, collision frequency increases with both the speed of the particles and the density of the mixture [6].

Similarly, spatial patterns of movement influence the probability of interactions in living systems, from bacterial aggregations, in which quorum sensing depends on chemical cues associated with local density [7], [8], to human online social networks [9], workplace interactions [10], and the votes of legislators [11]. In animal groups, spatial proximity influences the spread of innovative foraging strategies [12], vocal communication [13], and fission-fusion group dynamics [14]. Proximity in turn may be determined by social attraction and preferential associations, or simply clustered resources [15], [16]. For simple interactions such as an alarm signal [17], proximity strongly determines the probability of interaction. However, some animals preferentially form stable associations with certain others, for example in the grooming interactions of baboons, that influence the probability that particular individuals interact [18]–[20].

In ant colonies, local interactions lead to collective behavior and decentralized decision-making [21]. Because ants interact locally, how they move around determines the rate of encounter [22]. Ants adjust their movement depending on the density of ants in the surrounding area [23]–[25]. Local interactions among ants are olfactory, such as the detection by one ant of a pheromone recently deposited by another [26] or brief antennal contact [21]. During an antennal contact one ant detects the other ant's cuticular hydrocarbon profile; such profiles differ among colonies and among task groups within a colony [27]. Workers engaged in different tasks organize spatially in different areas of the nest [28]. Ant nests are diverse in structure [29]–[31]. The spatial configuration of the nest [32], [33] and the spatial organization of task groups affects the rates at which ants interact [34], so that nest structure influences colony behavior [35].Changes in nest structure as a colony develops may further influence colony behavior [21], [36], [37].

Here we examine how spatial patterns of movement determine the interaction patterns that regulate foraging activity in the harvester ant, *Pogonomyrmex barbatus*. Whether a forager waiting inside the nest, here called a "potential forager", leaves on its next foraging trip depends on its interactions with returning foragers [38]. Foragers with food return to the nest and enter an entrance chamber just inside the nest entrance, where they deposit the seeds they have collected. Potential foragers come into the entrance chamber from tunnels leading from the deeper nest. After some time in the entrance chamber, where they interact with returning foragers, they either leave the nest to forage, or return to the deeper nest. Previous work shows that potential foragers that leave the nest to forage have more interactions with returning foragers than those that leave the nest to forage [39]–[41].

We use models from statistical mechanics to examine how the spatial patterns of movement determine the probability of interaction of returning and outgoing harvester ant foragers inside the nest. Our analysis is based on data from observations of returning and outgoing foragers inside nests of freely foraging colonies of *Pogonomyrmex barbatus* in the field [41]. Laboratory and field studies show that harvester ants form interaction hotspots with a high local density of ants in the entrance chamber or its analogue in a laboratory nest [40], [42]. We first examine the temporal autocorrelation of the time series of interactions and the average density of surrounding ants to ask how an ant's experience of interactions is structured in time. If ants mix homogeneously and have a constant relationship between proximity and interaction, then interactions are described by a Poisson process with a flat autocorrelation function. If engaging in an interaction leads an ant to seek future interactions, then the interaction autocorrelation function will be peaked near zero. However, a heterogeneous ant density will also lead to a peak in the interaction autocorrelation by causing interactions to be clustered in time. By comparing the interaction autocorrelation with the interaction-triggered density average, we ask whether an ant's experience deviates from a constant rate Poisson process. We then examine whether this is this due to facilitation, when an ant that has an interaction seeks another, or to density hotspots.

Next we ask whether returning and potential foragers mix homogeneously in the entrance chamber, or instead differ in their use of space. We use a correlated random walk model to ask whether the observed heterogeneity in density can be explained by the different starting locations of returning and potential foragers.

We then use collision theory to formalize the relationship between walking speed, ant density, and interaction, and derive two models with increasing complexity: (1) a random mixture model, assuming homogeneous mixing in the entrance chamber, and (2) a local density model, in which the probability of interaction depends on walking speed and local density of surrounding ants. The local density model is conceptually similar to a proximity network [20], which assumes that animals interact when in close proximity. We use the local density model to distinguish among several possibilities for what determines the probability that returning and potential foragers interact:
  (1) Returning foragers (RF's) and potential foragers (PF's) preferentially interact with each other. If true, then observed interaction rates would be higher than the local density model's predictions, because RF-PF interactions are more likely than RF- or PF- interactions with other ants. The local density model assumes all ants have the same relationship between spatial proximity and the probability of interaction.
  (2) Returning foragers and potential foragers preferentially avoid interactions with each other. This is the opposite of (1), and would lead to observed interaction rates that are lower than local density model predictions.
  (3) When potential foragers become excited and prepare to leave the nest to forage, they seek interactions with returning foragers, and avoid interactions with other potential foragers. If this were true, potential foragers that leave the nest to forage ("PF-leave") would have a higher probability of interaction than potential foragers that returned to the deeper nest ("PF-return"). In this case, the local density model, which uses the same relationship between density and the probability of interaction for all ants, would underestimate the interaction rates of PF-leave.
  (4) There is individual variability in walking speed, but no systematic differences between potential foragers that leave the nest to forage, and those that return to the deeper nest, in the relationship between density and interaction rates with returning foragers.

# Methods

## Field experiments and video tracking

We used data on the movement and interactions of harvester ants in the entrance chamber of actively foraging colonies, from observations from videos made in the field inside partially excavated nests [41]. Interactions of potential foragers with returning foragers take place inside the nest entrance chamber, and tunnels lead from this chamber to the deeper nest. Films of the entrance chamber were made by removing the top layer of soil above the entrance chamber and placing a transparent piece of glass over it to maintain humidity [39]–[41].

We manually tracked the trajectories and interactions of approximately 1200 foragers in the entrance chamber during a tracking period of 60-180 seconds for each observation. An interaction was considered to occur when the ant's head came within one head width of another ant; this is about the length of the antennae in this species. Fig 1 shows the entrance chamber for each colony, and all the trajectories and interactions that were tracked during this period.

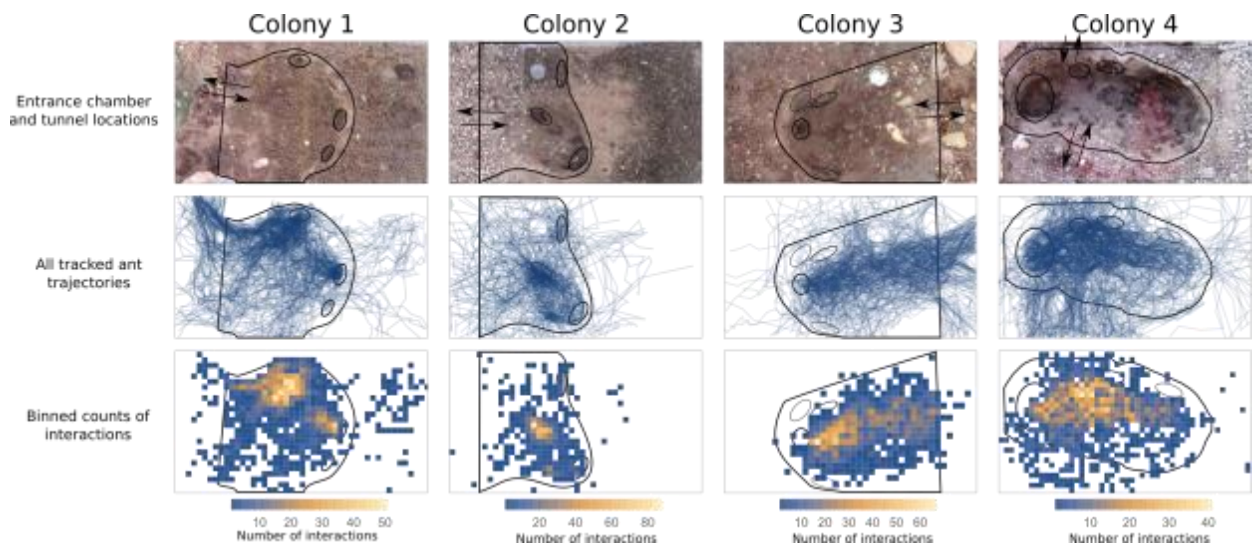

**Figure 1. Tracked trajectories and locations of interactions of all ants.** (Top row) The entrance chamber of each observed colony, showing the entrance chamber boundary and tunnel locations. Arrows indicate the most commonly used directions of entry and exit. For colonies 1-2, ants tended to enter and exit the entrance chamber from the left side of the video frames; for colony 3, from the right side, and for colony 4, from the top and bottom. (Middle row) An overlay of all trajectories during the observation period. (Bottom row) A histogram of the location of interactions.

We categorized each ant as either a returning or potential forager according to the place where its trajectory began. Returning foragers came back into the entrance chamber from outside the nest, while potential foragers came into the entrance chamber from one of the tunnels leading from the deeper nest. Once inside the entrance chamber, both returning foragers and potential foragers exited after a short time (ca. 10-30 sec). Each was classified as taking one of two actions: leave the nest to forage, or return to the deeper nest through a tunnel [41]. For ants that were in the entrance chamber when the tracking period began, we identified their start location from earlier video to classify them as returning or potential foragers. Further details on methods are included in supplemental material, as well as in [39]–[41].

1. Density hotspots and the temporal sequence of interactions

The temporal autocorrelation of the time series of interactions was calculated for all tracked ants in each observation. This calculation uses a window time of $\tau$ and selects interactions for each focal ant that occur up to $\tau$ seconds before or $\tau$ seconds after a given interaction. Interactions that occurred within the first or last $\tau$ seconds of a focal ant's trajectory were not included in the average. We then binned these counts of interactions, neglecting each count at $t = 0$, to obtain an autocorrelation of interaction rate. The window time of $\tau = 4$ sec. was used for the results in the text.

The temporal autocorrelation was compared with the average density of ants surrounding a focal ant in the time before and after each interaction. To do this, we smoothed the locations of ants to form a density function. Each ant was represented as a 2D Gaussian function centered at the $(x, y)$ coordinates of its current location and a standard deviation $\sigma = 2d_{ant}$, where $d_{ant}$ is the size of an ant in a colony video observation. This representation blurs an ant over an area larger than its actual size, and thus enables a measure of the local density of other ants near a given focal ant. To facilitate calculations, the 2D Gaussian was approximated by three concentric circles of radii $\sigma$, $2\sigma$, and $3\sigma$, each given weights of $cg_1$, $cg_2$, and $cg_3$, respectively. The values $g_1 = 0.683$, $g_2 = 0.272$, and $g_3 = 0.043$ are calculated from the Gaussian probability distribution, and the constant $c$ ensures normalization via $c\pi\sigma^2(g_1 + 3g_2 + 5g_3) = 1$. Fig 2A shows an example of this representation of a single ant. Fig. 2B shows an example of the time-dependent density function for all ants, $\rho_D(x, y, t)$, at a particular instant in time.

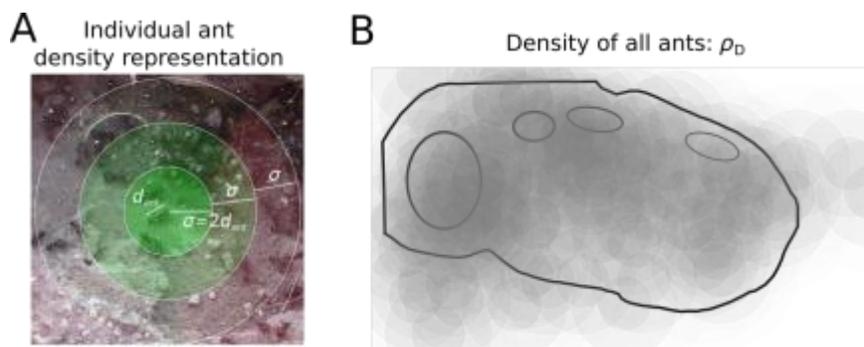

**Figure 2. Ant density function.** (A) A single ant with the approximate 2D density function overlaid. (B) Example density function for all tracked ants ($\rho_D$) at a snapshot in time for colony observation 4.

For a focal ant $i$ with trajectory coordinates $(x_i(t), y_i(t))$, the density of all other ants is $(\rho_D(x_i(t), y_i(t), t) - c\pi\sigma^2 g_1)$, where the correction factor subtracts the focal ant's own contribution to density. This corrected density value was used to calculate an average density in the time range $t_{int} - \tau \leq t \leq t_{int} + \tau$ surrounding an interaction at $t_{int}$. Interactions of all tracked ants, other than those that occurred within the first $\tau$ seconds or the last $\tau$ seconds of a focal ant's trajectory, were used to trigger the density average. To plot the average density along with the temporal autocorrelation on the time series of interactions, a normalization factor was used:

$$c_D = \frac{\sum_i N_i}{\sum_i \int_{T_i^{start}}^{T_i^{end}} (\rho_D(x_i(t), y_i(t), t) - c\pi\sigma^2 g_1) \, dt}$$

where $N_i$ is the number of interactions of ant $i$, and the sums include all tracked ants. Using $c_D$ for normalization, the normalized density $c_D\rho_D$ has units of interaction rate and was plotted alongside the temporal autocorrelation.

An additional comparison was made with the expected flat autocorrelation function from a Poisson process, where the average rate was defined as the overall average rate of interaction observed in each colony.

## 2. Density and mixing of potential and returning foragers

We first compared the walking speeds and densities of returning and potential foragers in the entrance chamber. Next we examined the effect of different start locations for returning foragers, who come into the entrance chamber from outside the nest, and for potential foragers, who come into the entrance chamber from a tunnel from the deeper nest. Using a null motion model, we considered whether heterogeneity of density and the existence of density hotspots in the entrance chamber could arise from the different starting locations of returning and potential foragers.

## 2.1 Walking speed of returning and potential foragers

Speed was calculated for each ant by finding the difference between successive tracked location points, smoothing these differences with a Gaussian kernel with a radius of 10 time steps, and then taking the average over the trajectory of an individual ant $i$ to obtain its average walking speed $\bar{s}_i$. Let the notation $\langle \cdot \rangle_{i \in G}$ denote an average for ants in group $G$. Only ants with completed trajectories, that both entered and exited the entrance chamber during the focus tracking period, were included in the group averages. We used a permutation test to ask if there was a significant difference between average group speeds for returning foragers, $\langle \bar{s}_i \rangle_{i \in RF}$, and potential foragers, $\langle \bar{s}_i \rangle_{i \in PF}$, for each colony observation.

## 2.2 Differences in density of returning and potential foragers

We asked if returning and potential foragers differ in their spatial organization in the entrance chamber by comparing the local density of returning and potential foragers surrounding each ant. To do this, we formed density functions using the trajectories of all tracked returning and potential foragers. Each individual ant's position was smoothed into a density function using the representation shown in Fig 2A. We denote the density function for all returning foragers as $\rho_D^{RF}(x,y,t)$, and for all potential foragers $\rho_D^{PF}(x,y,t)$. An example of these density functions at a particular instant in time is shown in Fig 3.

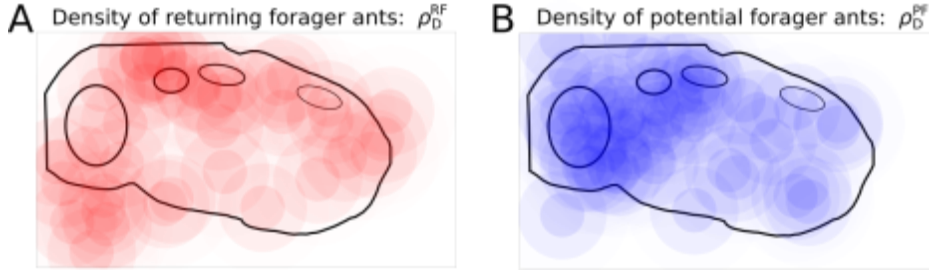

**Figure 3. Density functions for returning and potential foragers.** Density functions at a snapshot in time for colony observation 4, for (A) Returning forager ants ($\rho_D^{RF}$), and (B) Potential forager ants ($\rho_D^{PF}$).

Consider an ant $i$ with time-dependent trajectory coordinates $(x_i(t), y_i(t))$, trajectory start time $T_i^{start}$, and trajectory end time $T_i^{end}$. For this ant, the average density difference between surrounding returning foragers and potential foragers over the course of its trajectory is calculated as

$$\overline{\Delta}_i = \frac{1}{T_i^{end} - T_i^{start}} \int_{T_i^{start}}^{T_i^{end}} [\rho_D^{RF}(x_i(t), y_i(t), t) - \rho_D^{PF}(x_i(t), y_i(t), t) - c\pi\sigma^2 g_1(\delta_{i,RF} - \delta_{i,PF})] dt$$

where the $\delta$-functions act to remove the focal ant from the associated density function. A permutation test was used to test for significance between average group density differences $\langle\overline{\Delta}_i\rangle_{i \in RF}$ and $\langle\overline{\Delta}_i\rangle_{i \in PF}$ for each colony observation. As for the speed averages described above, only ants with completed trajectories, that both entered and exited the entrance chamber during the focus tracking period, were included in the group averages.

## 2.3 Density differences: spatial constraints or selective attraction?
To ask whether density differences between returning and potential foragers could be explained by their different start locations, we used a null model which represents each ant as a correlated random walk. A simulated set of trajectories corresponding to a focal ant $i$ were generated as a walker with constant speed of $\overline{s}_i$, beginning at the point $(x_i(T_i^{start}), y_i(T_i^{start}))$. The turning dynamics follow an Ornstein-Uhlenbeck process [43]:

$$d\omega = -\alpha dt + \sigma dW(t)$$

where $\omega$ is the turning rate, $\alpha$ is the inverse correlation time, $\sigma$ is the noise amplitude, and $W(t)$ is a Wiener process. The parameters $\alpha = 7$ sec$^{-1}$ and $\sigma = 5$ rad/sec were chosen as representative of the ant trajectories. A manual search suggested that the statistical comparison results were not sensitive to specific parameter choices. An additional simulation rule was used to keep the simulated trajectories inside the entrance chamber: if the next simulation step would place the trajectory outside the entrance chamber, then instead a new motion direction that maintains the trajectory inside the chamber was chosen at random. A set of 50 correlated random walk trajectories were simulated for each tracked ant, each having a total simulation time of $T_i^{end} - T_i^{start}$. Let the set of simulated trajectories for ant $i$ be $\{(x_i^*, y_i^*)\}$. The average density difference for these simulated trajectories is

$$\overline{\Delta_i^*} = avg\left\{\frac{1}{T_i^{end} - T_i^{start}}\int_{T_i^{start}}^{T_i^{end}}[\rho_D^{RF}(x_i^*(t), y_i^*(t), t) - \rho_D^{PF}(x_i^*(t), y_i^*(t), t)]\,dt\right\}$$

where the average is over the ensemble of of 50 simulated trajectories, and the $\delta$-function correction is not used here because the trajectories are simulated and therefore not contained in the density functions. A permutation test was used to test if there were significant differences in $\langle\overline{\Delta_i^*}\rangle_{i\in RF}$ and $\langle\overline{\Delta_i^*}\rangle_{i\in PF}$ for each colony observation. Figure 4A shows an example of a simulated correlated random walk using the start location and total walking time of an ant from colony 4, and Figure 4B shows the ensemble of simulated trajectories generated for this ant.

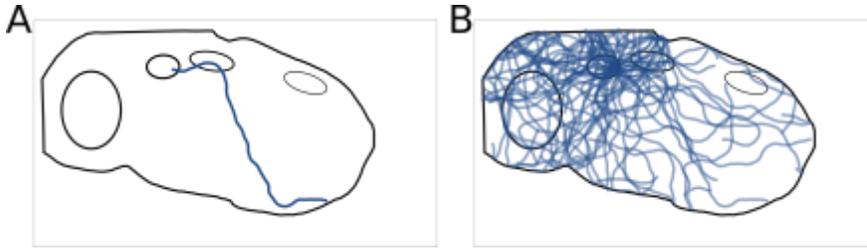

**Figure 4. An ant as a correlated random walk.** (A) A single simulated trajectory with starting location and walk time corresponding to a potential forager ant from colony 4. (B) A set of 50 simulated trajectories with the same starting location and walk time as (A).

### 3. Effect of density on rate of interactions between returning and potential foragers

We used collision theory to compare two groups of potential foragers based on their foraging decision, those that left the nest to forage or those that returned to the deeper nest. We asked whether differences between the two groups in interaction rates are consistent with chance encounters between ants. We then compare potential foragers in two quantities that influence their interaction rate: walking speed, and the density of surrounding returning foragers.

### 3.1 Collision theory model of interactions between returning and potential foragers

We applied collision theory to predict each potential forager's expected rate of interaction with returning foragers. It is expected that the encounter rate should increase with both walking speed and local density of surrounding ants. To formalize this in terms of collision theory, consider a focus particle with size $d_{ant}$ and speed $s$, surrounded by a density of particles $\rho$. The expected collision rate per unit time is $f = d_{ant}s\rho$. However, since this calculation assumes that the other particles are stationary, it underestimates the expected collision rate. Using a mean field approximation, the expected collision rate considering the relative velocity of a focal particle $i$ with respect to neighboring particles $j$ is approximated with the following expression:

$$f = d_{ant}\sqrt{(s_i\rho)^2 + \langle s_j\rho\rangle_j^2}$$

where $\langle s_j\rho\rangle_j$ is an average of speed times density of local neighbors $j$. See supplemental for a derivation of this expression.

To assess how individual walking speed and local density affect a potential forager's probability of encountering returning foragers, we consider two models of increasing complexity:

1) **Random mixture model**: Individual potential foragers walk at different speeds, and experience a uniform density of returning foragers throughout the entrance chamber.
2) **Local density model**: Individual potential foragers walk at different speeds, and experience a changing local density landscape defined by the returning forager density function $\rho_D^{RF}(x, y, t)$.

For the random mixture model, the density function is simply $M_{RF}(t)/A$ for each ant, where $M_{RF}(t)$ is the number of returning foragers in the entrance chamber as a function of time and $A$ is the area of the entrance chamber. The expected collision rate for a focal ant $i$ is

$$f_{mix,i}(t) = d_{ant} \frac{M_{RF}(t)}{A} \sqrt{s_i(t)^2 + \langle s_j(t)\rangle_{j\in RF}^2}$$

where $\langle s_j(t)\rangle_{j\in RF}$ is the average speed of all returning foragers in the entrance chamber at time $t$, and the subscript $mix$ represents the random mixture model.

For the local density model, the expected collision rate for ant $i$ is

$$f_{LD,i}(t) = d_{ant} \sqrt{\left(s_i(t)\rho_D^{RF}(x_i(t), y_i(t), t)\right)^2 + \langle Q_{ij}(t)\rangle_{j\in RF}^2}$$

where the subscript $LD$ stands for local density. Here, $\langle Q_{ij}(t)\rangle_{j\in RF}$, the local average of speed times density of returning foragers $j$ surrounding a focal ant $i$, is calculated using the same approximate 2D Gaussian function described above:

$$\langle Q_{ij}(t)\rangle_{j\in RF} = c\left(g_1 \sum_{\substack{j,\\ d_{ij}\leq\sigma}} s_j(t) + g_2 \sum_{\substack{j,\\ \sigma<d_{ij}\leq 2\sigma}} s_j(t) + g_3 \sum_{\substack{j,\\ 2\sigma<d_{ij}\leq 3\sigma}} s_j(t)\right)$$

where $d_{ij}$ is the distance between ants $i$ and $j$.

For both the random mixture and local density models, the expected collision rate is used to express a probability function:

$$P_{mix,i}(t) = \frac{f_{mix,i}(t)}{Z_{mix}}$$

$$P_{LD,i}(t) = \frac{f_{LD,i}(t)}{Z_{LD}}$$

where the associated partition functions are

$$Z_{mix} = \sum_{i=all\ ants} \int_{T_i^{start}}^{T_i^{end}} f_{mix,i}(t)\, dt$$

$$Z_{LD} = \sum_{i=\text{all ants}} \int_{T_i^{start}}^{T_i^{end}} f_{LD,i}(t)\, dt$$

The average expected number of interactions for an ant is obtained by integrating over its trajectory using the above probability functions. We denote the average number of expected interactions for ant $i$, calculated for either model, as $\overline{N_{mix,i}}$ or $\overline{N_{LD,i}}$. The actual count for the number of interactions that ant $i$ made with returning foragers is $N_i^{RF}$. For normalization we require that the total number of observed interactions be equal to the total average expected interactions for each model:

$$N^{RF} \equiv \sum_i N_i^{RF} = \sum_i \overline{N_{mix,i}} = \sum_i \overline{N_{LD,i}}$$

For an individual ant, the average expected number of interactions with returning foragers is then calculated as follows:

$$\overline{N_{mix,i}} = N^{RF} \int_{T_i^{start}}^{T_i^{end}} P_{mix,i}(t)\, dt$$

$$\overline{N_{LD,i}} = N^{RF} \int_{T_i^{start}}^{T_i^{end}} P_{LD,i}(t)\, dt$$

This calculation can be considered to distribute the interactions among all of the observed ants according to each model's probability function, and yields the average of the model's predicted number of interactions for each ant.

For each potential forager ant $i$ that completed a trajectory during the tracking period, we plotted the observed rate of interaction with returning foragers, $r_i = N_i^{RF}/T_i$, versus the average predicted from each model, $\overline{r_{mix,i}} = \overline{N_{mix,i}}/T_i$ or $\overline{r_{LD,i}} = \overline{N_{LD,i}}/T_i$, where $T_i$ is the ant's total time in the entrance chamber. The correlation between model and data was calculated by grouping together potential foragers for all 4 sets of observations.

Previous work showed that potential foragers that left the nest to forage tended to interact with returning foragers at a higher rate than potential foragers that returned to the deeper nest [41]. We tested the ability of the random mixture and local density models to capture trends in the interaction rates of the two groups of potential foragers, those that left the nest to forage and those that returned to the deeper nest, by comparing predicted and observed mean group interaction rates. In the data, the mean rate of interaction with returning foragers is $\langle r_i \rangle_{i \in PF-leave}$ for potential foragers that left the nest to forage, and $\langle r_i \rangle_{i \in PF-return}$ for potential foragers that returned to the deeper nest. The standard error of the mean was calculated via bootstrapping. For the models, the mean predicted rates are $\langle \overline{r_{mix,i}} \rangle_{i \in PF-leave}$, $\langle \overline{r_{mix,i}} \rangle_{i \in PF-return}$, $\langle \overline{r_{LD,i}} \rangle_{i \in PF-leave}$, and $\langle \overline{r_{LD,i}} \rangle_{i \in PF-return}$. An additional simulation step is used to represent the discreteness of interactions and obtain distributions of these model-predicted group mean quantities. At each time step during an ant's trajectory, a random number is drawn between 0 and 1. If this number is less than the probability function times $N^{RF}$, an interaction occurs at this timestep, otherwise there is no interaction. This condition is written as $rand[0,1] < N^{RF} P_{mix,i}(t)$ or $rand[0,1] < N^{RF} P_{LD,i}(t)$. This process is applied to each model to obtain distributions of the possible numbers of interactions that are consistent with the probability function. Fig 5A illustrates the collision rate calculation, and Fig 5B shows an example of time-dependent interaction probabilities $N^{RF}P_{mix,i}(t)$, and $N^{RF}P_{LD,i}(t)$ for a potential forager.

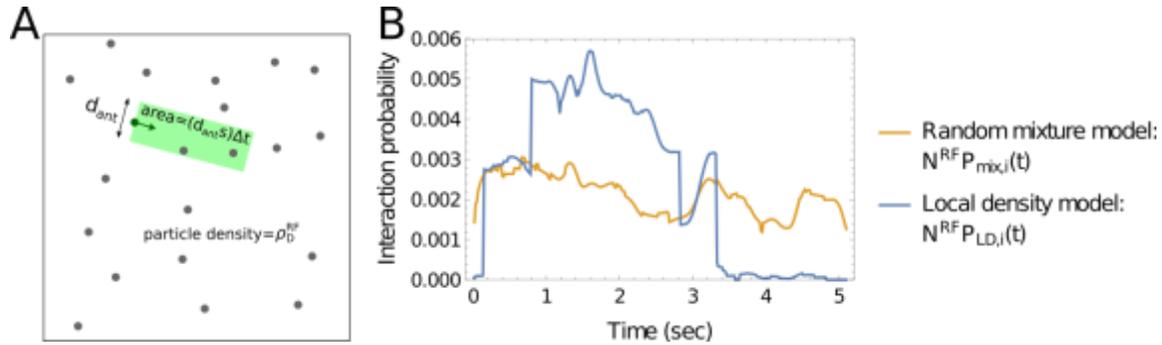

**Figure 5. Collision theory illustration and example time-dependent interaction probabilities.** (A) Collision theory states that a particle with speed $s$ and size $d_{ant}$ will sweep out an area of $(d_{ant}s)\Delta t$ in a short time $\Delta t$. In an environment with particle density $\rho_D^{RF}$, a focal particle will encounter other particles at a rate of approximately $d_{ant}s\rho_D^{RF}$. (B) An example of the time-dependent probability that a potential forager will interact with a returning forager, calculated with both collision theory models, the random mixture model and the local density model.

### 3.2 Walking speed and surrounding returning forager density, based on foraging decision

We compared the average walking speed and the average proximity to returning foragers, for potential foragers that left the nest to forage and potential foragers that returned to the deeper nest. The calculation for an individual's average walking speed ($\overline{s_i}$) is described in Section 2.1. A permutation test was used to test for a significance difference between the group average speeds $\langle\overline{s_i}\rangle_{i \in PF-leave}$ and $\langle\overline{s_i}\rangle_{i \in PF-return}$.

Proximity to returning foragers was quantified by averaging the density function for returning foragers, $\rho_D^{RF}(x,y,t)$, over the trajectory of each focal potential forager ant $i$:

$$\overline{D_i^{RF}} = \frac{1}{T_i^{end} - T_i^{start}} \int_{T_i^{start}}^{T_i^{end}} \rho_D^{RF}(x_i(t), y_i(t), t)\, dt$$

A permutation test was used to test for a significant difference between group average densities $\langle\overline{D_i^{RF}}\rangle_{i \in PF-leave}$ and $\langle\overline{D_i^{RF}}\rangle_{i \in PF-return}$. Only ants with completed trajectories, that both entered and exited the entrance chamber during the tracking period, were included in the group averages.

# Results

## 1. Density hotspots and the temporal sequence of interactions

We find that the temporal autocorrelation for each observation displays a broad increase near zero, and that the trend is well matched by the interaction-triggered density average (Fig 6). This result has a simple interpretation: ants engage in more interactions in high density areas in the entrance chamber, and this causes an ant's experience of interactions to be clustered in time. The match between the interaction autocorrelation and the density average suggests that engaging in an interaction does not lead an ant to seek further interactions. Instead, density hotspots cause an ant's experience of interactions to be clustered in time and to deviate from a constant rate Poisson process.

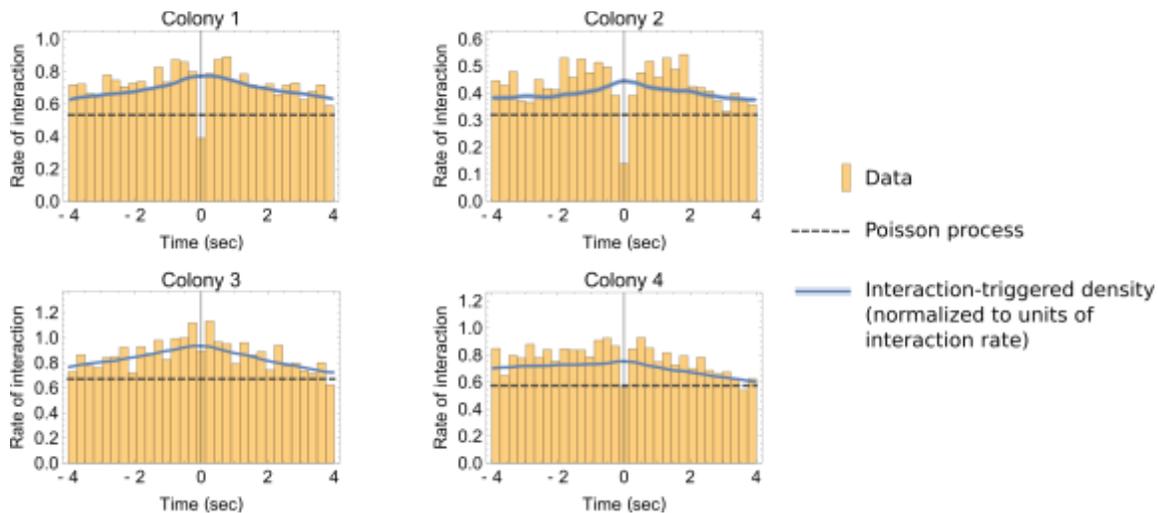

**Figure 6. Temporal autocorrelation from data, compared to a Poisson process and the normalized density average.** The temporal autocorrelation of interactions calculated from the data (binned histograms) is compared with a Poisson process with the same average rate as each observation (dashed line), and with the interaction-triggered average of local density (solid and shaded lines, showing mean and standard error).

## 2. Density and mixing of potential and returning foragers

### 2.1 Walking speed of potential and returning foragers

The mean walking speed of returning foragers was higher than the mean walking speed of potential foragers in each observation. However, the differences in the mean walking speed between groups of ants were not large, and were only marginally significant for colonies 1 and 4 (colony 1, p=0.046; colony 2, p=0.12; colony 3, p=0.13; colony 4, p=0.046).

### 2.2 Differences in density of potential and returning foragers

Returning foragers tended to spend proportionally more time near returning foragers, and potential foragers tended to spend proportionally more time near other potential foragers. In observations of colonies 1, 3, and 4, the average density difference for returning foragers was significantly higher than the average density difference for potential foragers (colony 1, p=0.006; colony 3, p=0.008; colony 4, p<0.001). In colony observation 2, there was no difference between returning and potential foragers (colony 2, p=0.80). The density differences tended to be negative because in each observation, there were fewer returning foragers than potential foragers. Fig 7A shows the distributions of individual walking speed and density difference for returning foragers and potential foragers, and Fig 7B shows the trajectories of returning forager and potential forager ants in colony observation.

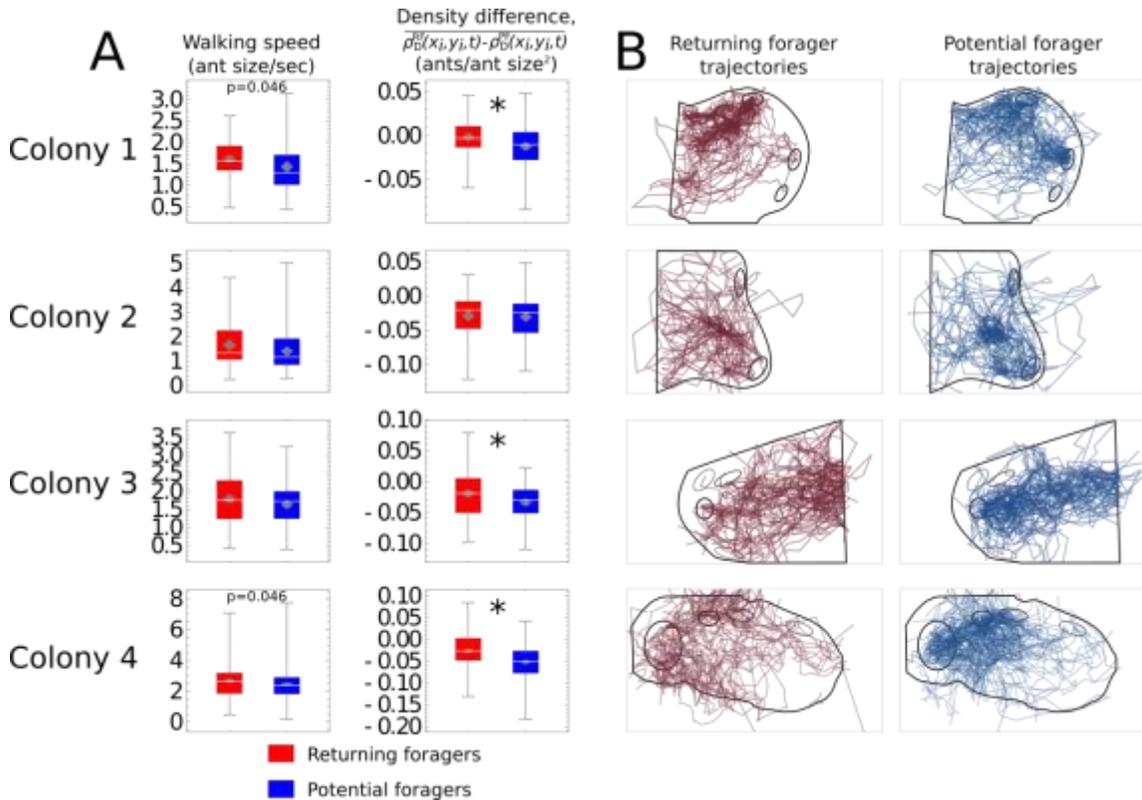

**Figure 7. Returning foragers compared to potential foragers: walking speed, surrounding density difference, and entrance chamber locations.** (A) Grouped comparison of returning foragers and potential foragers, showing box-and-whisker plots for average individual walking speed and average individual surrounding density difference of returning foragers minus potential foragers ($\overline{\rho_D^{RF}(x_\iota, y_\iota t) - \rho_D^{PF}(x_\iota, y_\iota t)}$). The diamond shows the confidence interval for the mean of each distribution. A significant difference in the means is denoted with an asterisk (permutation test, significance at 0.05 level). (B) Grouped ant trajectories for all returning foragers (left column) and all potential foragers (right column). The results in each row are identified by the corresponding colony label.

## 2.3 Density differences: spatial constraints or selective attraction?

Using a null motion model, we considered whether the tendency of returning and potential foragers to spend time in different locations in the entrance chamber was associated with starting location. Returning foragers came into the entrance chamber from outside, and potential foragers came into the entrance chamber from a tunnel to the deeper nest. In the null motion model, each simulated trajectory begins at the same starting location as the corresponding tracked ant, but then moves randomly. The results of the null model analysis show that differences in the locations of returning and potential foragers in the entrance chamber can indeed be explained by their starting locations. The simulated trajectories predict a significant density difference between groups of returning forager and potential forager ants (Fig 8A).

We additionally compare visually the simulated null motion model trajectories with the observations. Although the null motion model trajectories show segregation in space due to different starting locations (Fig 8B), the simulated trajectories do not display as prominent of hotspot areas as in the observations. This is particularly noticeable for the simulated and observed trajectories of returning and potential foragers in colony observations 1 and 3, and of potential foragers in colony observation 4. The visual comparison of simulated and observed trajectories suggests that to accurately reproduce realistic hotspot areas with a motion model would require more realistic detail in entrance chamber structure as well as traffic constraints between ants.

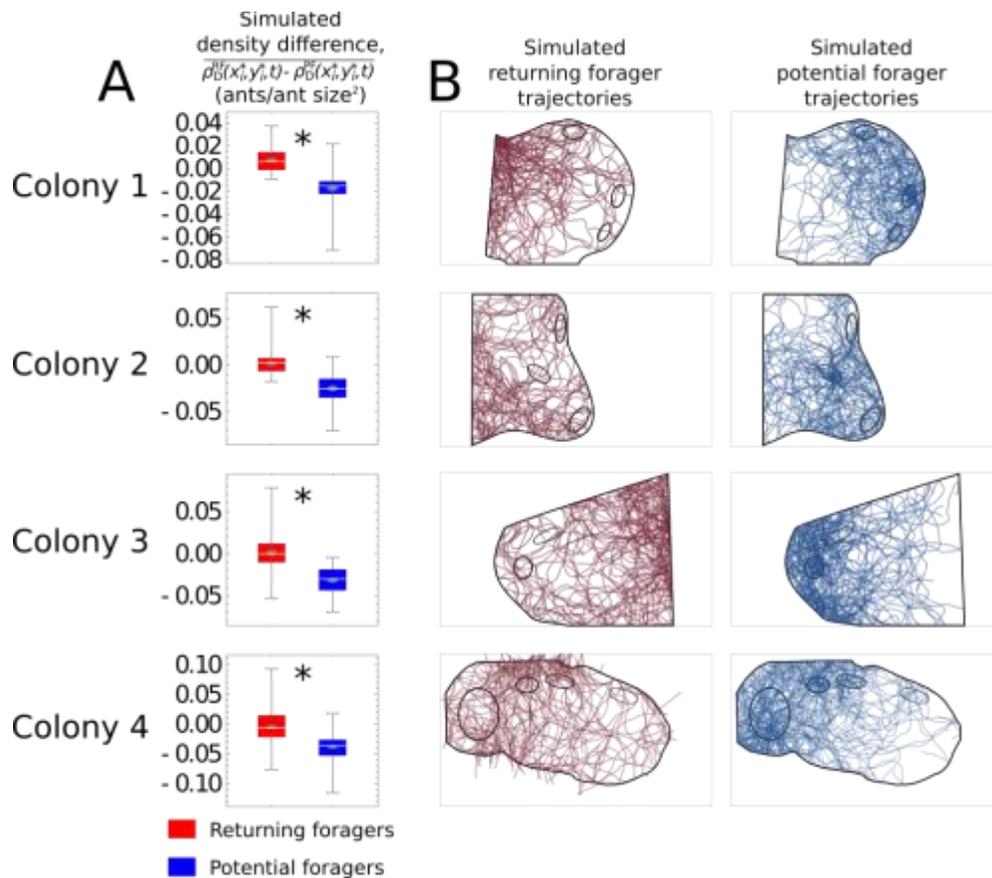

**Figure 8. Simulated density differences and trajectories using the correlated random walk model.** Each forager begins with the same starting location as in the tracked data, but then moves as a correlated random walk with no other movement preference. (A) Density differences calculated using simulated trajectories for returning and potential foragers. (B) Simulated ant trajectories corresponding to returning foragers (left column) and potential foragers (right column). One simulated trajectory is shown for each ant.

## 3. Effect of density on rate of interactions between potential and returning foragers
### 3.1 Collision theory models applied to interactions between potential and returning foragers

The results of comparing the collision theory model with the observations are consistent with the conclusion that there are no systematic differences in the relationship between density and interaction rate for the two potential forager decision outcomes, either leave the nest to forage, or return to the deeper nest. First, the local density model agrees well with the data for the average rate of interaction of potential foragers with returning foragers (Fig 9A). This suggests that potential foragers neither seek nor avoid interactions with potential foragers, but instead interact with surrounding returning foragers at the same rate as other surrounding ants. Next, the local density model shows a general agreement with the data for the difference in interaction rate between potential foragers that left the nest to forage (PF-leave) compared to potential foragers that returned to the deeper nest (PF-return) (Fig 9B). In each case, the local density predicts that PF-leave interact on average at a higher rate than PF-return. For colony observations 1 and 3, the agreement between model and data is very close. For colony observations 2 and 4, the model prediction is slightly below the observations, but the predicted distributions still overlap with the error range for the data.

The random mixture and local density models make nearly the same average predictions in some cases, for example the average rate of interaction of potential foragers in colonies 1-3, and the difference in interaction rates of potential foragers in colonies 2 and 4. However, in other cases, such as the difference in interaction rates of potential foragers in colonies 1 and 3, the predictions of the two models differ. This may be due to differences among colonies in the spatial structure and distribution of ants in the entrance chamber (Fig 7).

The correlation between model and data for all potential foragers is shown in Fig 9C. Although the local density model improves the correlation with the data over the random mixture model and matches the group averages well, there is still considerable variation among individual ants. In particular, the local density model predictions do not match well for potential foragers that interacted at a high rate with returning foragers, or potential foragers that interacted with no returning foragers (Fig 9C).

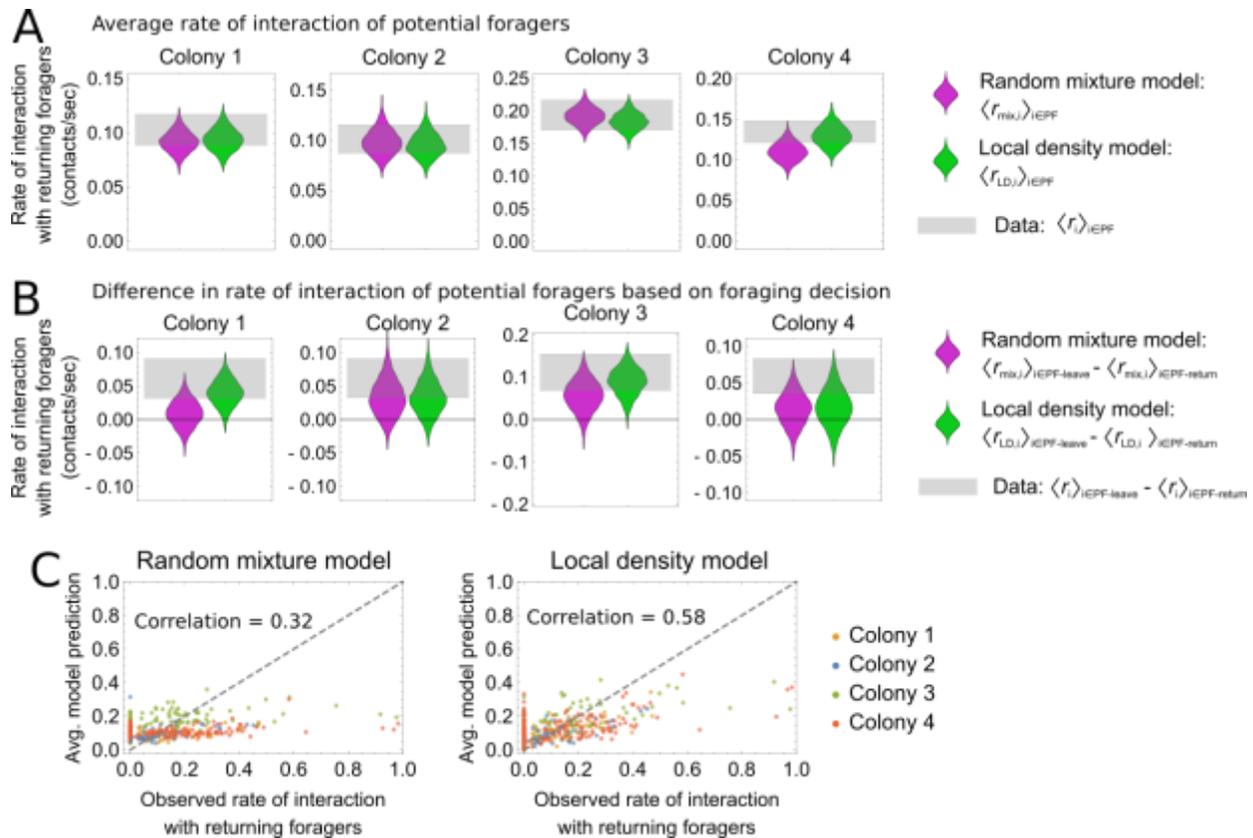

**Figure 9. Collision theory models applied to predict the rate of interaction of potential foragers with returning foragers.** (A) Average interaction rate of potential foragers with returning foragers, showing model predictions and the observed result. The blobs show the distributions obtained from each model. The shaded gray line shows the mean and standard error of the mean from the data. (B) Difference in interaction rates between potential foragers that left the nest to forage (PF-leave) and potential foragers that returned to the deeper nest (PF-return). Model and data representation are the same as (A). (C) Correlation between model and data. For both the random mixture and local density models, the average model prediction for each potential forager is plotted as a function of the observed rate of interaction with returning foragers. Points are colored by colony observation.

### 3.2 Walking speed and surrounding returning forager density, based on foraging decision

In all observations, the average walking speed of potential foragers that left the nest to forage was significantly higher than that of potential foragers that returned to the deeper nest (colony 1, p=0.005; colony 2, p<0.001; colony 3, p<0.001; colony 4, p<0.001). Although potential foragers that left the nest to forage engaged in interactions with returning foragers at a higher rate [41], there were no consistent trends or significant differences in the average density of returning foragers in the area surrounding potential foragers that took either action (colony 1, p=0.37; colony 2, p=0.94; colony 3, p=0.37; colony 4, p=0. 48). The distributions are shown in Fig 10A.

All potential foragers emerged from a tunnel and therefore spent time near tunnel areas while in the entrance chamber. Potential foragers that left the nest to forage walked near the edges of the entrance chamber, while potential foragers that returned to the deeper nest rarely did (Fig 10B). However, this did not cause ants that left the nest to forage to spend more time near returning foragers (Fig 10A).

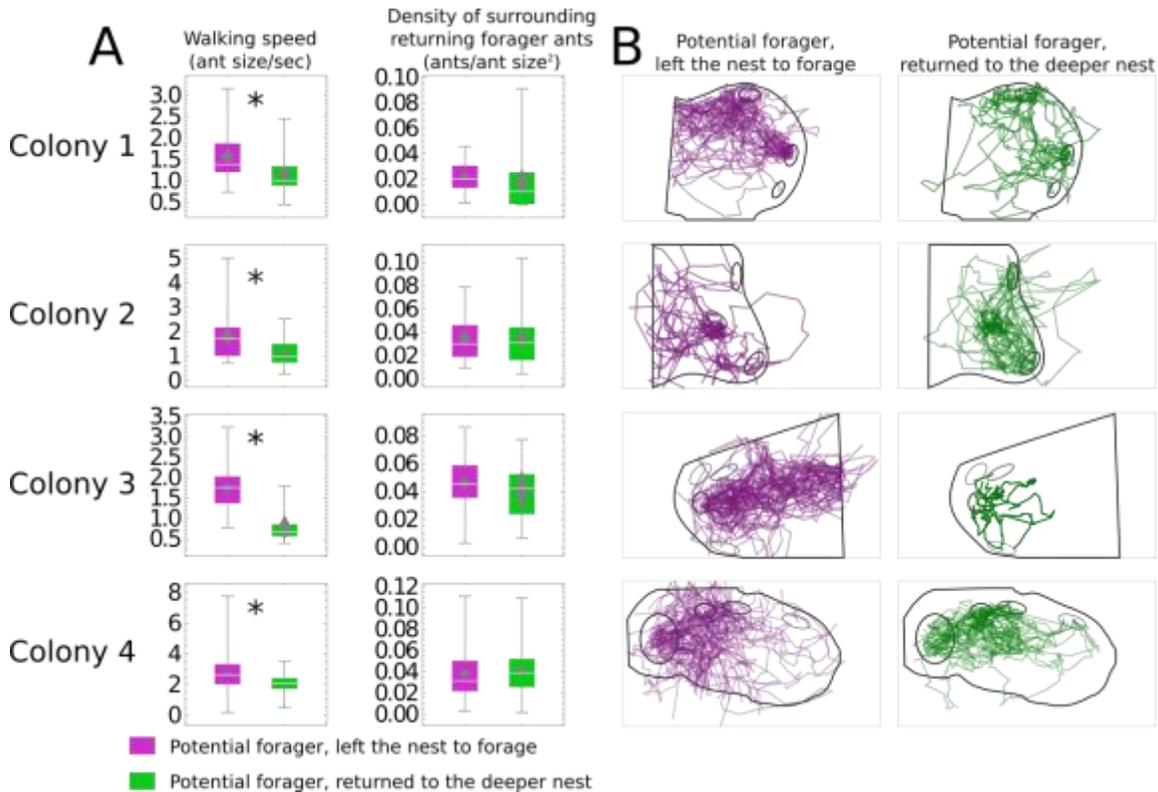

Figure 10. Potential foragers that left the nest to forage compared to potential foragers that returned to the deeper nest: walking speed, surrounding density of returning foragers, and entrance chamber locations. (A) Grouped comparison of potential foragers that left the nest to forage, versus potential foragers that returned to the deeper nest. Shown are box-and-whisker plots for average individual walking speed and average individual surrounding density of returning foragers ($\overline{\rho_D^{RF}(x_\iota, y_\iota, t)}$). The diamond shows the confidence interval for the mean of each distribution. A significant difference in the means is denoted with an asterisk (permutation test, significance at 0.05 level). (B) Grouped ant trajectories for all potential foragers that left the nest to forage (left column) and all potential foragers that returned to the deeper nest (right column). The results in each row are identified by the corresponding colony label.

## Discussion

We formulated a model to relate proximity and walking speed to the probability of interaction between potential and returning foragers (Figs 5, 9). Building on this, we asked, do some ants 'seek' interactions more so than others, or does engaging in an interaction make an ant more likely to seek other interactions? Examining the temporal autocorrelation (Fig 6) shows that
17

although interactions occur clustered in time due to density hotspots, engaging in an interaction does not change the relationship between proximity and the probability of interaction Instead, the density hotspots in the entrance chamber cause ants to experience interactions that are clustered in time. Although returning and potential foragers do not mix homogeneously in the entrance chamber (Fig 7), differences in their use of space are simply due to different starting locations (Fig 8). Potential foragers that left the nest to forage walked faster and interacted with returning foragers at a higher rate than potential foragers that, instead of going out to forage, returned to the deeper nest (Fig 10).

The collision theory results for group average interaction rate (Fig 9A) suggest that there are no systematic differences between returning foragers and potential foragers in interaction preferences, or between potential foragers that left the nest to forage and those that returned to the deeper nest. Returning foragers do not selectively seek potential foragers to interact with, or vice versa. However, the density hotspots in the entrance chamber suggest that ants do tend to adjust their motion towards other ants. In species that form preferential associations between particular individuals, such as birds and primates, spatial proximity does not always correspond strongly with interactions [18], [20]. In harvester ants, however, spatial proximity corresponds to the probability of interactions.

We found that potential foragers differed in their walking speed, depending on whether they subsequently left the nest to forage or returned to the deeper nest. Previous work has suggested that interactions with returning foragers cause potential foragers to increase their walking speed as they become excited and subsequently leave the nest to forage [44]. Our measurements of walking speed from field observations are not precise enough to ask whether potential foragers increased their speed immediately after an interaction with a returning forager, as was noted in a laboratory study with a different species of harvester ants [44]. Another possibility is that ants differ in their preferred walking speed, or in sensitivity to interactions and the subsequent decision whether to leave the nest to forage [41].

Returning and potential foragers differ in their use of space in the entrance chamber because of their different starting locations: Returning foragers were more likely to be around other returning foragers, and potential foragers more likely to be around other potential foragers. In ants, the effect of spatial structure on task allocation and task performance is likely to have an important effect on colony organization in many species [28], [34].

Although collision theory captures the group average differences in interaction rate, there was still considerable variation among individual ants in the observed interaction rates (Fig 9C). Each ant was observed in the entrance chamber for a short time before it left to forage or entered a tunnel to other parts of the nest; longer observations might smooth out this variation. However, individual ants probably vary in their movement patterns, for example in how they turn in response to the movement of nearby ants. However, it is clear from the random mixture model that some of the observed differences among ants can be explained solely by walking speed, and the fit between model and data is further improved by considering the local density of returning foragers surrounding each potential forager ant.

The collision theory model establishes a baseline expected rate of interactions based on proximity which could be used to examine interactions in other social groups. This is a null model for the case when there is no social preference. In some animals, such as baboons [18] and giraffes [45], individuals are more likely to interact with social affiliates than with other group members. Comparing the model's baseline expectation of interactions with observations of animal groups can help to elucidate how social preference and spatial constraints from the environment influence the interactions that shape collective behavior.

## Data accessibility
All data is freely available in the Stanford data repository at https://purl.stanford.edu/sf994km1767. Code used in the analysis was written in Mathematica and is available at https://github.com/jacobdavidson/ants-spatial.

## Acknowledgements
We are grateful to Mark Goldman for helpful discussions that initiated this work, and to Paul Switzer for discussions and statistics advice. We thank Sam Crow and Roxana Arauco-Aliaga for help with field work and ant tracking, as well as the staff at Southwestern Research Station for their help with field work. This work was supported by NIH R01 GM105024.

# Supplemental Methods

Field experiments were performed with colonies of the red harvester ant *Pogonomyrmex barbatus* at the site of a long term study near Rodeo, NM, USA, monitoring a population of about 300 colonies for which the ages of all colonies are known (Gordon and Kulig 1996; Ingram et al. 2013). Observations were made in August 2013 and August 2014.

Interactions of potential foragers with returning foragers take place inside the nest entrance chamber, a chamber about 5 cm long, and 2-3 cm below the surface. The entrance chamber connects to the nest entrance by a small tunnel, and further tunnels lead from this chamber to deeper nest areas. Films of the entrance chamber of four actively foraging colonies were made by removing the top layer of soil above the entrance chamber and placing a transparent piece of glass over it to maintain humidity (Pinter-Wollman et al. 2013; Davidson et al. 2016; Pless et al. 2015).

We manually tracked the interactions and locations of all ants in 1 to 3 minutes of each video using a Java program we developed. An interaction was considered to occur when the tracked ant's head came within one head width of another ant. The location of an ant was marked with a tracking point when it significantly moved positions or changed the course of its trajectory in a subsequent video frame; this allowed an approximate reconstruction of the ant's entire trajectory by linearly interpolating between tracking points. Ants that were in the entrance chamber when the focus period began were followed back in time to establish if they were returning foragers or had come from a tunnel.

Trajectories were classified based on starting and end location, whether the ant was seen carrying objects, and whether the ant left a tunnel to walk around the entrance chamber. The classification of ant trajectories used here is that same as (Davidson et al. 2016), and includes the trajectory categories showing in Table S1.

The density function $\rho_D^{RF}$ for returning foragers includes ants with labels {f, g, or h} in Table S1, and the density function $\rho_D^{PF}$ for potential foragers includes ants with labels {a, b, or d}. These labels are included in the linked data files.

To compare groups of returning foragers and potential foragers (Fig 7), we calculated averages using only individual ant trajectories that were completed during the focus period. For returning foragers, this included categories {f, g}, and for potential foragers categories {a, b}, both with the restriction of the ant completing its trajectory during the focus tracking period. The additional label of whether the ant completed its trajectory during the focus period is included in the linked data files. Similarly, the comparison of potential foragers ants (Figs 8-9) uses labels a or b, with the condition of the ant completing its trajectory during the focus period.

| Label | Description | Notes |
|---|---|---|
| a | From tunnel, left to forage | |
| b | From tunnel, into tunnel | |
| c1 | From tunnel, carrying (not a potential forager) | Emerged from a tunnel carrying an object, and thus were likely to have been engaged in in nest maintenance work. |
| c2 | From tunnel, always in tunnel (not a potential forager) | Never left a tunnel area to walk around the entrance chamber |
| c3 | From tunnel, down tunnel, shorter than fastest outgoing forager (not a potential forager) | Returned to the deeper nest but stayed in the entrance chamber for a time less than that of the fastest outgoing forager |
| d | From tunnel, lost | Lost during tracking |
| e | From tunnel, uncertain action (colony 1) | Colony 1 (field study colony #242) had a location in the upper right area of the video frame that was an area of active nest maintenance. Ants that emerged from a tunnel that were not seen to be carrying dirt, but ended their trajectory near this area were labeled as having an uncertain end action. |
| f | Returning forager, left to forage | |
| g | Returning forager, into tunnel | |
| h | Returning forager, lost, uncertain, or incomplete trajectory | This includes ants that were lost during tracking, not tracked until trajectory completion (for ants that appeared near the end of the focus tracking period), or ended their trajectory in the upper right area of the video frame for colony 1 (an area of active nest maintenance) but were otherwise not seen to be carrying dirt. |
| i | Not many tracking points | Had 3 or fewer trajectory location makers placed during tracking. These ants either appeared in the video frame for a very short time, often at the edge of the frame or never leaving a tunnel area, or were lost and could not be followed any further |
| j | Other (uncertain start action) | These do not fall into any of the other categories, and mainly include ants that could not be followed back in time to discern where they started from, or ants that entered the video frame but never entered the entrance chamber. |
| k | Nest maintenance | If an ant carried dirt or debris out of the nest, it was considered to be a nest maintenance worker. For colony 1, if an ant first appeared in the upper right area of the video frame, it was also considered to be a nest maintenance worker since the colony was engaged in maintenance work in this area. |

**Table S2. Labels and descriptions of categories of tracked ants.**

## Approximation for average relative speed between ants

Here we approximate the average speed between a focus ant $i$ and the surrounding ants $j$. Let the velocity of the focus ant be $\vec{v_i}$, and the surrounding ants $\vec{v_j}$. The average relative speed is an average over the surrounding ants:

$$\langle s_{rel,i} \rangle = \left\langle \sqrt{(\vec{v_i} - \vec{v_j}) \cdot (\vec{v_i} - \vec{v_j})} \right\rangle_j$$

Assume that the velocities of the focus ant and the surround ants are uncorrelated. We then then neglect the term containing $\vec{v_i} \cdot \vec{v_j}$, and approximate the relative speed as

$$\langle s_{rel,i} \rangle \approx \sqrt{s_i^2 - \langle s_j \rangle_j^2}$$

where $\vec{v_i} \cdot \vec{v_i} = s_i^2$, and a mean field approximation is also used to move the averaging over $j$ inside the square root. This expression is used to evaluate the expected collision rate for the random mixture model.

For an average over local density and relative speed, we wish to compute the quantity

$$\langle \rho_D s_{rel,i} \rangle = \left\langle \rho_D \sqrt{(\vec{v_i} - \vec{v_j}) \cdot (\vec{v_i} - \vec{v_j})} \right\rangle_j$$

Again assumed uncorrelated velocities and using a mean field approximation, we arrive at

$$\langle \rho_D(x_i(t), y_i(t), t) s_{rel,i}(t) \rangle \approx \sqrt{\left(s_i(t) \rho_D(x_i(t), y_i(t), t)\right)^2 - \left\langle \left(s_j(t) \rho_D(x_i(t), y_i(t), t)\right)^2 \right\rangle_j}$$

Here we note explicitly that the density function is evaluated at the current location of the focus ant $i$, and that the speed of each ant is a function of time. This expression is used to evaluate the expected collision rate for the local density model.